# COVID-19 Blockchain Framework: Innovative Approach

Mohamed Torky[1,*] and Aboul Ella Hassanien[2,*]

[1]Faculty of Computer and Information Systems, Islamic University in Madinah, KSA
[2]Faculty of Computers and Artificial Intelligence, Cairo University, Egypt

[*]Scientific Research Group in Egypt (SRGE), Egypt

**Abstract:** The world is currently witnessing dangerous shifts in the epidemic of emerging SARS-CoV-2, the causative agent of (COVID-19) coronavirus. The infection, and death numbers reported by World Health Organization (WHO) about this epidemic forecasts an increasing threats to the lives of people and the economics of countries. The greatest challenge that most governments are currently suffering from is the lack of a precise mechanism to detect unknown infected cases and predict the infection risk of COVID-19 virus. In response to mitigate this challenge, this study proposes a novel innovative approach for mitigating big challenges of (COVID-19) coronavirus propagation and contagion. This study propose a blockchain-based framework which investigate the possibility of utilizing peer-to peer, time stamping, and decentralized storage advantages of blockchain to build a new system for verifying and detecting the unknown infected cases of COVID-19 virus. Moreover, the proposed framework will enable the citizens to predict the infection risk of COVID-19 virus within conglomerates of people or within public places through a novel design of P2P-Mobile Application. The proposed approach is forecasted to produce an effective system able to support governments, health authorities, and citizens to take critical decision regarding the infection detection, infection prediction, and infection avoidance. The framework is currently being developed and implemented as a new system consists of four components, Infection Verifier Subsystem, Blockchain platform, P2P-Mobile Application, and Mass-Surveillance System. This four components work together for detecting the unknown infected cases and predicting and estimating the infection Risk of Corona Virus (COVID-19).

# 1. Introduction

The confirmed SARS-Cov-2, the causative agent of Wuhan new corona virus disease 2019 (COVID-19) cases exceeded SARS-CoV-1, the causative agent of severe acute respiratory syndrome (SARS) cases. At the time of writing this paper, there are currently +740,201 confirmed cases +28,000 critical/serious cases and +35,026 deaths worldwide from the COVID-19 pandemic outbreak as of March 30, 2020, 11:23 GMT **[1]**. Coronavirus is often presented with a novel word, as a new strain can be within the virus family, we have all discovered beforehand. As indicated by the World Health Organization (WHO), Coronaviruses belong to a large family that varies from cold to serious disease **[2]**.

These diseases can affect humans and animals. The strain that began to spread in Wuhan, the capital of China's Hubei region, was identified from two completely different Coronaviruses namely, severe acute respiratory syndrome (SARS) and the Middle East respiratory syndrome (MERS). Symptoms of coronary virus infection increase the severity of respiratory complications such as respiratory disorder, kidney disorder and fluid development in the lungs. The entire healthcare organization will eventually need to adjust to blockchain technology applications to support and fight COVID-19 outbreaks. In a globalized world in which we live, entities such as the World Health Organization have put in place the World Health Organization to ensure cooperation between different governments on issues related to global health. In the face of epidemics such as the one we are witnessing at the moment, these organizations need to obtain correct and accurate information to judge the best course of action: as we have observed over the past three months, quickly, reliable scientific data, sharing and reporting of this data, is the key to understanding, thus reducing global outbreaks. Besides, how the infected cases can be automatically detected and how the probability of the infection risk can be estimated and predicted by persons or authorities in real-time.

Blockchain technology application to the healthcare industry can improve information security management; healthcare data can be analyzed and transmitted while maintaining data privacy and security **[3]**. Here we take a critical look at how blockchain technology provide solutions about how the COVID-19 infected cases can be automatically detected and how the probability of the infection risk can be estimated and predicted by persons or authorities in the real times. By nature, we refer to the blockchain as a "decentralization" technique, which means that data is not stored in a single server but is shared among many participants, stores, and websites. As a result, health documents added to the blockchain can be accessed by all interested parties (who have the required permissions) in a transparent and reliable environment. This new technology has been proposed to disable a wide range of data-driven areas, including the health field.

In this COVID-19 epidemic, it is necessary to think of solutions to introduce blockchain as one of the emerging technologies as it guarantees health data security and privacy and provide appropriate tools for patient assessment, screening, testing, treatment and exchange of infection prevention and control information with patients and the public; as needed, appropriate security measures for personal safety are provided. For example, by giving each patient a unique identifier and the creation of an intelligent contract between patients and professionals that ensures that the joint data is reliable and accurate and thus will increase efficiency, with a faster diagnosis. Also, due to the inherent transparency of technology, hospitals can ensure that all patients' health records are fake, while data and privacy are still guaranteed.

Creating this COVID-19–blockchain platform will facilitate the interoperability of registry sharing between healthcare stakeholders with important information available to them in a non-mediated and efficient manner while ensuring patient privacy and security. Most importantly, cooperation between health workers around the world could also be enhanced. Accessing a patient's record can be a time consuming exercise and may delay patient care. It may also lead to poor management of these records and, in the worst case scenario, misdiagnosis. Digitizing records with a blockchain can reduce these shortcomings by creating a network where reliable information is immediately available.

This paper proposes a COVID-19 Blockchain platform as a new approach could be applied to fight the COVID-19 diseases outbreaks by automatically detecting unknown infected cases as well as predicting and estimating the contagion risk of COVID-19 epidemic in the real-time for communities. Peer-to-peer blockchain, time stamping, and decentralized storage features can support and strengthen the proposed system with important features in case of infection detection, contagion risk prediction and estimation of COVID-19 virus in real societies.
The proposed platform can be designed based on four subsystems, infection verification subsystem, a Blockchain platform, P2P-Mobile application, and a Mass-Surveillance system. These subsystems are discussed in more details in the next section.

## 2. The Proposed COVID-19 Blockchain Framework

Recently, several techniques are increasingly introduced for predicting COVID-19 growth and spread around the world. Some studies utilized Adaptive Neuro-Fuzzy Inference system (ANFIS) as a common approach that integrate fuzzy logic and neural networks in time series prediction and forecasting problems for forecasting COVID-19 growth **[4].** Other studies used further time series prediction models for forecasting other epidemics such as human West Nile virus (WNV) **[5]** , Hepatitis **[6]**, seasonal outbreaks of influenza **[7,8].** Moreover, additional mathematical models have been proposed for forecasting other epidemics such as Ebola **[9]**, SARS **[10]**, H1N1-2009 **[11],** and MERS **[12]**.

All these studies can be simply used also for forecasting COVID-19 growth with only modifying the inputs of those algorithms to be for COVID-19 growth forecasting. However, these techniques didn't introduce any solutions about how the infected cases can be automatically detected and how the probability of the infection risk can be estimated and predicted by persons or authorities in the real times. All these approaches focused on the prediction and growth curves of the number of confirmed cases, death cases, and recovered cases in next periods without real investigation for these critical challenges we mentioned.

In response to these important challenges, we try in this study to introduce a novel blockchain-based system for automatically detecting the infected cases and estimating the infection risk of COVID-19 epidemic in the society's real time. The peer-to peer , time stamping, and decentralized storage advantages of blockchain can support and strength the proposed system with significant features in detecting the infected cases and estimating the infection risk of COVID-19 epidemic in the societies real times. The proposed system can be designed based on four subsystems, Infection Verifier Subsystem, Blockchain platform, P2P-Mobile Application, and **Mass-Surveillance System**. The general architectural model of the proposed system is described in **Figure 1.** The infection Verifier Subsystem is responsible for digitally representing, creating, verifying infection patterns and infection instances. *Blockchain platform* work as a decentralized repository for storing all data about confirmed COVID-19 cases in a sequence of digital blocks in real-time. P2P Mobile Application system can be used by citizens and authorities for visualizing important information about infection risk estimation results , statistics, prediction data and detecting infected cases based on P2P communication between citizens and authorities. Mass-Surveillance System is a complex surveillance system of citizens able to monitor the contacts between citizens and tracking motions of specific persons for detecting the contact persons and places of a confirmed COVID-19 case has contacted with.

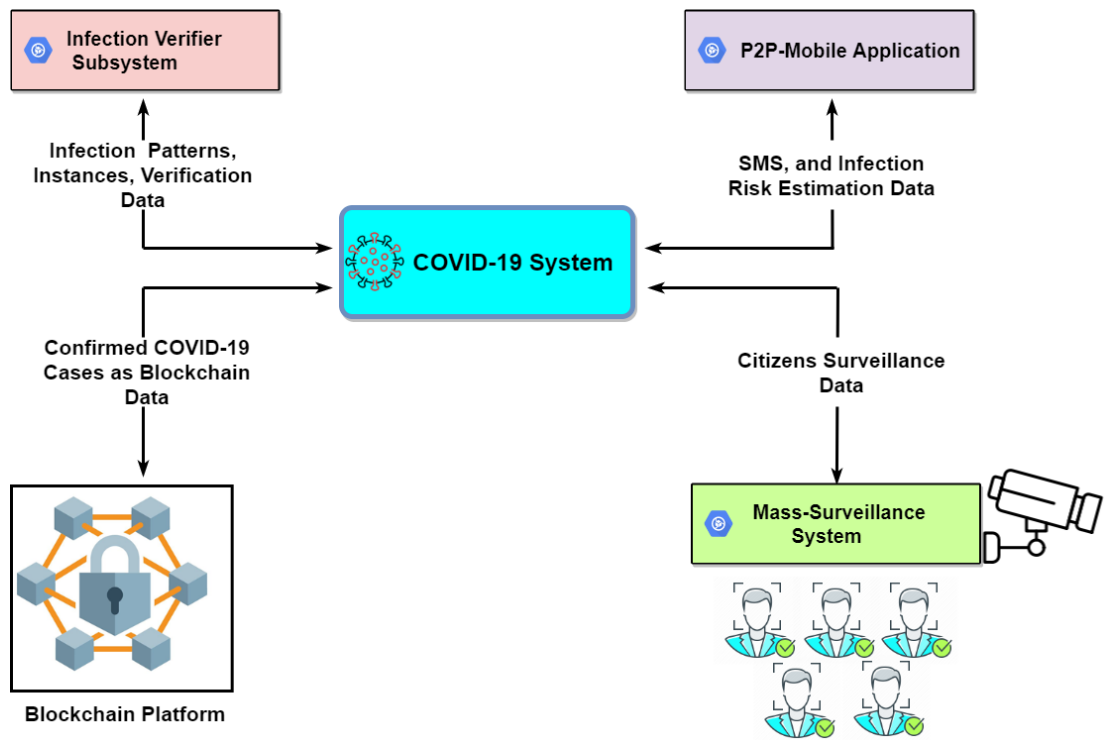

Figure 1: Architectural Model of the proposed COVID-19 System: A Proposed Model

## 2.1 Infection Verifier Subsystem

The first confirmed case that holds the COVID-19 virus is the seed of the infection spread. Hence, the follower infected cases can be generated by the close contact with the first contagion source (i.e. the seed). The way by which the contagion source and his/her infected cases are digitally represented can be simulated using *Regular Expression approach* **[13].** Regular expression (far away specifying programming languages as we know about the common utilization of Regular Expression) is an interesting technique to represent data (or object) patterns and its derived follower instances. The main advantage of using regular expression for representing data patterns is the wide range of instance-derivation possibilities, you can create one regular expression pattern and derive so many follower instances to this pattern, and it is a very good motivation to Specify COVID-19 Cases using Regular Expression Patterns.

So, imagine that, we can digitally represent the contagion source of Covid-19 using a regular expression when it is confirmed as contagion source using the appropriate medical tests, we call it the *infection Pattern*. The set of persons who the contagion source made a close contact with, and the set of places which the contagion source went to can be digitally represented with the derived instances from the corresponding pattern, which we call it *infection instances*. So, each *infection pattern* can generates many *infection instances*.

For example, in **Figure 2**, the regular expression, $P|Bab^+c$ can be used to represent the *infection pattern* of a confirmed case of Covid-19. This pattern can generates many *infection instances* which digitally represent the infected *persons* and infected *buildings* in the form of "***Pabc***", "***Pabbc***", "***Pabbbc***", "***Pabbbbc***",…., ***Babc***, "***Babbc***", "***Babbbc***", "***Babbbbc***",…etc. The infection instances that start with '***P***' represent *infected persons* whilst the infection instances that start with '***B***' represent *infected buildings*, such as schools, homes, supermarkets, parks, and other public places which always have many people and may be risky places if infected persons have been entered.

Verifying the *infection instances* to confirm the infection transferee to a specific person can be achieved by a *Finite Automaton Model* which it is known that it is the best recognizer and verifier technique for patterns (i.e. regular expressions) **[14]**. So, for each *infection pattern*, there is a unique Finite Automata model which is responsible for verifying all *infection instances* that follow that *infection pattern.* For example, if a person who has the infection instance "***Pabbbbc*** has been accepted by the corresponding Automaton model, this mean that this person has high probability of contagion

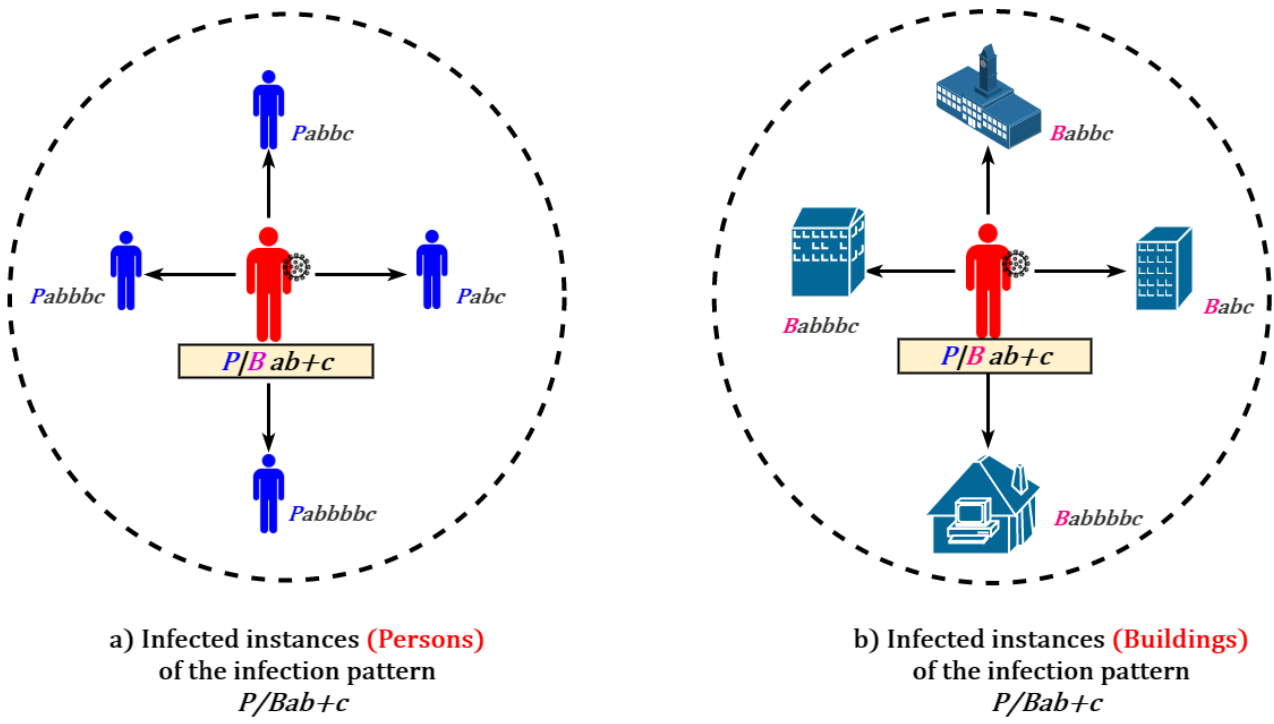

**Figure 2:** Using Regular Expression technique for specifying infection pattern and the corresponding infection instances for : a) infected persons, and b) infected buildings : A prosed Model

that has been transferred from a confirmed infected case that represented by the infection pattern, $P|Bab^+c$. By this way, the new infected persons (i.e. new infected instances) can know the source of the contagion. **Figure 3** explain how the infection verifier subsystem recognizes the *infected instances* that are corresponding to the infection *pattern*, $P|B\ 0+(0+1)^*0+1(0+1)^*$.

Additional benefits of using this technique for verifying COVID-19 contagion are that the *infection patterns* codes can be used to digitally store and track the confirmed COVID-19 cases in the blockchain as a new added block. On the other hand, the *infection instances* codes will be used in P2P communications between all citizens using the P2P-mobile application. The *infection instances* will be also used in estimating the contagion risk and infection probability between citizens. In the coming two sections, we will explain in more details how *blockchain* can be used as P2P repository and tracker for confirmed COVID-19 cases (section 3.2), and how the *P2P mobile Application* can be used for detecting the unknown infected cases by automatically sending SMS notifications between persons and the relevant authorities. In addition, the P2P Mobile App will be used for estimating contagion risk from infected persons and infected places to help people to take their precautions (section 3.3).

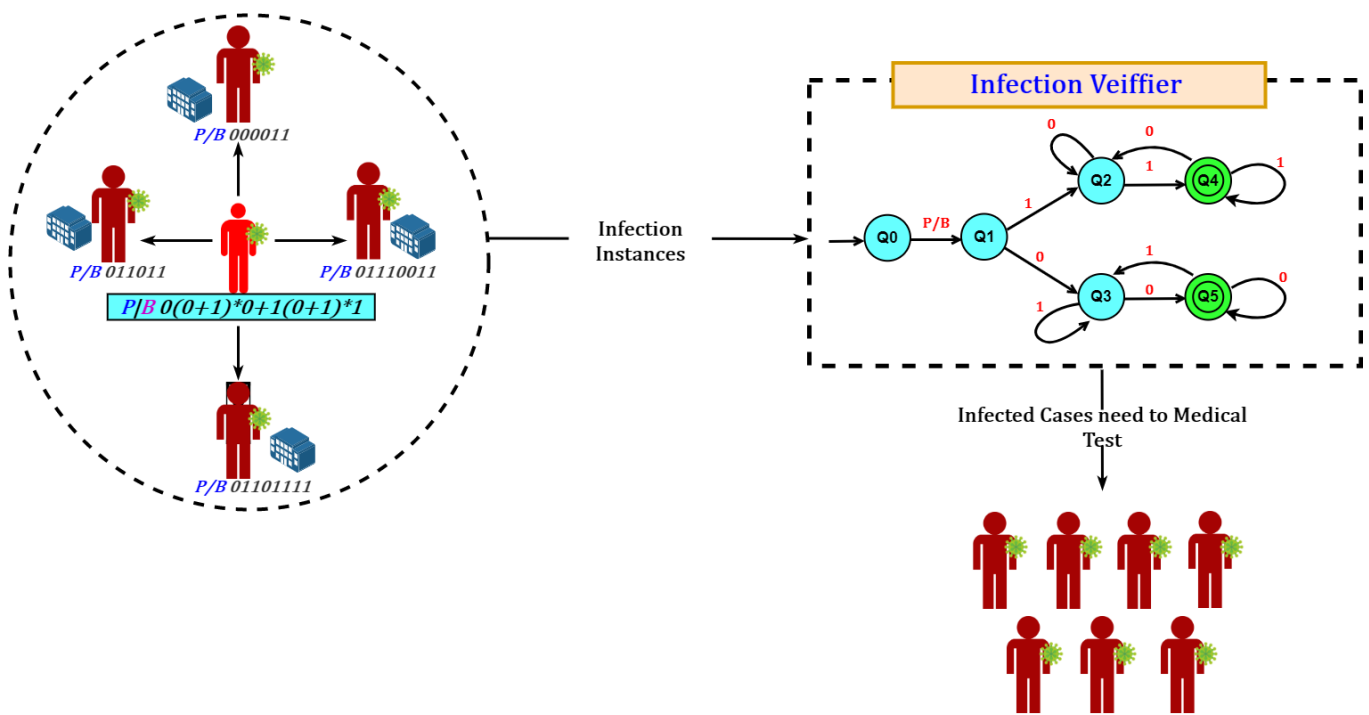

**Figure 3.** COVID-19-Infection Verifier Subsystem: A proposed Model

## 2.2 Blockchain Platform

The decentralized design, time stamping, and real time-tracing features of blockchain, make it is an effective solution for digitally storing all confirmed COVID-19 cases based on the associated *infection pattern* of each confirmed case. Storing a new confirmed case mean that a new corresponding infection *pattern* is created and stored, in addition, all contact persons and risky places which have been accessed by a confirmed COVID-19 case are also stored in the block chain as *infection instances* . The new stored *infection patterns* and the corresponding *infection instances* represent a new added block to blockchain at a specific time. Moreover the P2P design of blockchain can enable to automatically *detect the unknown infected cases*; this can be achieved by a hidden interaction that occurs between blockchain and the P2P mobile application once a new block is added to blockchain.

With the proposed system in this study, we propose an update to the block design as well as specifying a new method by which a new block is added to the blockchain. The classical and known design of blockchain which was used since the advent of digital currencies **[15]** is not effective manner with the proposed system's functionality. This is due to the difference in stored data types and the mining way by which the new block is validated and added to the blockchain. Hence, the two important questions are: what is the new update in the block design? And how a new block is added to the blockchain?

The new block design consist of two parts: The block header and the block body as depicted in the proposed design model in **Figure 4**. The block header consists of four components which can be categorized as follows:

1) *Block version:* is the way in which the data structures is organized inside the block and specifies how peers can read the block content using their computer or mobile APP. So in this component the data of *infection patterns* (i.e. Confirmed COVID-19 cases ) and the corresponding *infection instances* (i.e. probable infected cases) are formatted and organized
2) *Previous Block hash code:* is a hash value (256-bit hash code) which used as an identification number to the previous block, hence it used for establishing chronology connection between blocks in the blockchain.
3) *Markle Root:* is the most significant component in the block header. It is used to hash each pair of *infection patterns* using a hash tree to produce *unique fingerprint* for all inserted data in the block in the form of *markle root code*. This code is the used to represent and identify all data patterns inside the block through a short hash code (see **Figure 4**).

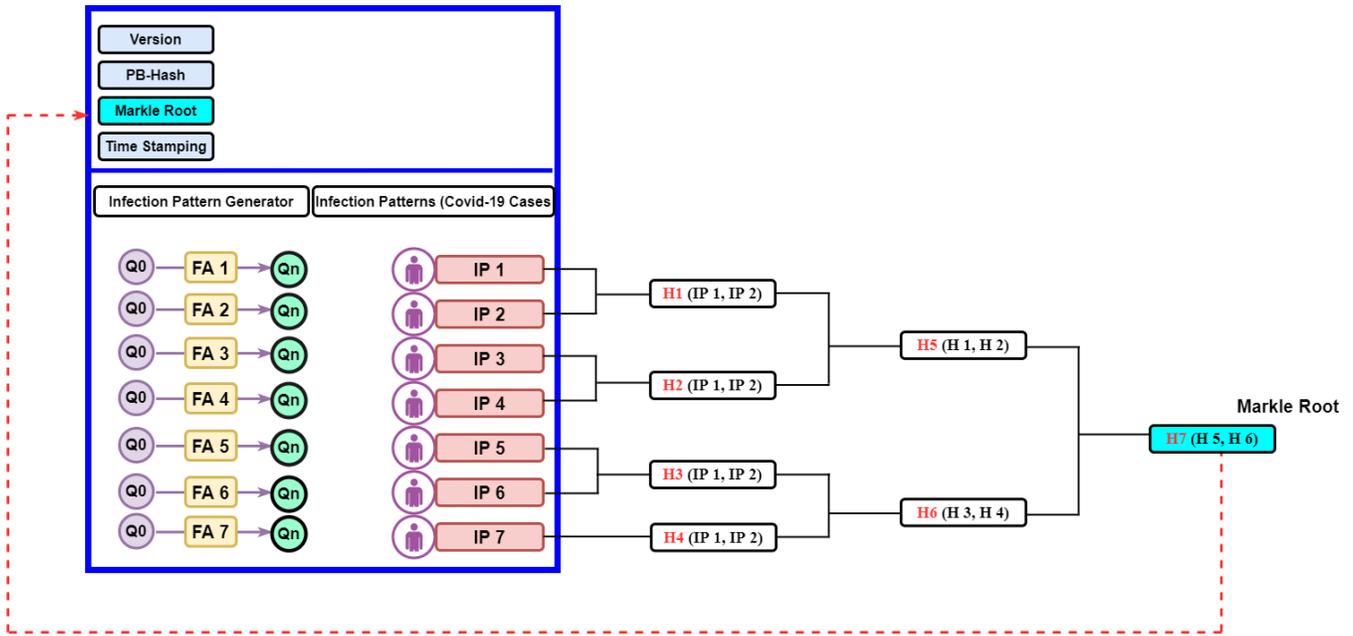

**Figure 4.** A proposed block design for processing Confirmed COVID-19 Cases: A Proposed Model

4) *Block Hash Code (BHC):* is the unique code that specifies the identity of the block. BHC can be calculated as in equation 1

$$BHC = Hash(MR + P + IS\ Code) \qquad (1)$$

Where, $BHC$ is the block hash code, $MR$ is the *markle rote* value, $P$ is the previous block's hash code, and $IS\ Code$ is a random value represents an infection *instance code*

5) *Time Stamping:* is a real time stamp given in the format DD/MM/YY-HH:MM:SS used to immutably track the creation and update time of the block for block integrity guarantee.

The block body consists of two components as follows:

1) *Infection Patterns of COVID-19 Cases :* is the space in which all detected confirmed C0VID-19 cases is digitized as *infection patterns* (IP) and stored in
2) *Infection Pattern Generator (IPG) :* we propose to add this component to the block' header. IPG is responsible for randomly generating a regular expression for each registered confirmed COVID-19 case in the form of a novel registered *infection pattern*. Moreover, IPG generate the corresponding *finite automata model* which is associated with the ID of the corresponding confirmed COVID-19 case then this verifier model is passed to the **infection verifier subsystem** for using it as a *verifier*

*model* for recognizing the relative infection *instances* (i.e. probable infected case ) . In addition, during the digital registration of a new confirmed COVID-19 case in the block, the IPG sends the identity information of the detected confirmed COVID-19 case to the ***Mass Surveillance system***, which , then transmits a feedback contains a list of all *contact persons* and places that this new patient has contacted or accessed ,then send it to the blockchain **(specifically for IPG)**. The IPG then specifies those contact persons and places and digitizes them as relative *infection instances* which then passed also to the P2P-Mobile Application, which then used for verifying the probable infected cases.

The second major question regarding blockchain utilization in the proposed system is how a new block is added to the blockchain, this process is called *block mining* **[15]**.

Each block contain a finite number of infection patterns (i.e. confirmed COVID-19 case) as depicted in figure 4. Adding new block can be described in the following steps:

**Step 0:** Suppose that the blockchain has only Block *0*, which contains 7 *infection patterns* (i.e. 7-confirmed COVID-19 cases). In addition, the follower *infection instances* (i.e. Probable infected cases) of this patterns which returned by ***Mass Surveillance system*** are sent to the **P2P-Mobile Application**, which is a mobile phone application with all citizens .The *infection instances* are propagated in the form of *SMS notification*.

**Step 1:** the SMS notifications are propagated only to persons who are identified as followers to the registered Confirmed COVID-19 cases (i.e. *infection patterns)* in the blockchain. This SMS message contains *a code* which called the *infection code*. This code will be used in two tasks. The first task is that, a specific person can log in the ***infection verifier subsystem*** and enter the *infection code* to verify his infection and know important information about the source of contagion, date of contagion, and the place of contagion, etc. The second task is that *the number of infection codes* which the person receives in the inbox of his P2P**-Mobile Application** can be used to estimate the probability of COVID-19 infection as we explain this procedure latter. **Figure 5** explains how parsons receive the infection codes from blockchain as SMS notifications and verify their contagion information through the ***infection verifier subsystem.*** The figure shows how the triple connections between *blockchain*, *P2P-mobile Application*, and *Infection verifier system* can detect three *probable infected groups* of parsons according to the corresponding registered confirmed COVID-19 cases in Blockchain.

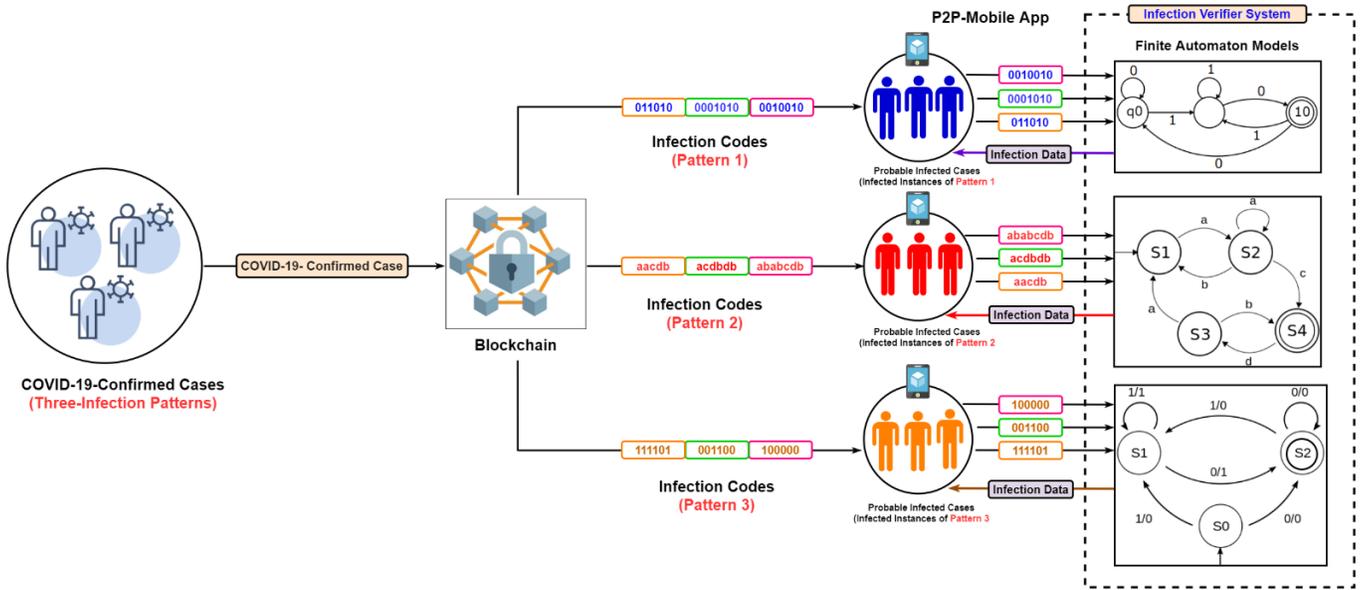

**Figure 5** A proposed models for identifying probable infected cases based triple connections between *Blokchain*, *P2P mobile Applications*, and *Infection Verifier system:* A proposed Model

**Step 2:** After detecting the number of *probable infected groups of persons* that are corresponding to the same number of registered Confirmed COVID-19 patterns in the blockchain, (e.g. Fig. 5 depicts three detected probable infected groups that are corresponding to three registered confirmed COVID-19 Patterns in Blockchain) the $N-$ number of all *infected instance codes* are distributed to $M-$number of pre-specified miners, (i.e. blockchain administrator staff), such that each miner will carry out set of tries to obtain the valid *Block hash code* (BHC) as formulated in **equation 1**.

**Step 3:** The set of miners try many times an *infection code* (one try for each code) for obtaining the valid hash code of that block. So every time a new *infection code* is tried, the block gets a new hash code. The Miners repeat this process by trying new *infection codes* many times until they randomly obtain the *infection code* that meets the target block hash code BHC. The new block is created when the valid BHC is obtained by the fastest miner. For example, if we have seven *infection patterns* registered in the blockchain as confirmed-COVID-19 cases, and the system detected corresponding 1000 *infected instances* (i.e.1000 persons who are probable infected cases) as well as the blockchain operator staff consist of four miners. So, each miner will try to obtain the valid hash code of the new block by doing repeated tries of the *infection codes* until the valid one that meets the target block hash code BHC is obtained. **Figure 6** explains how blockchain is constructed as sequence of new blocks according to the proposed design.

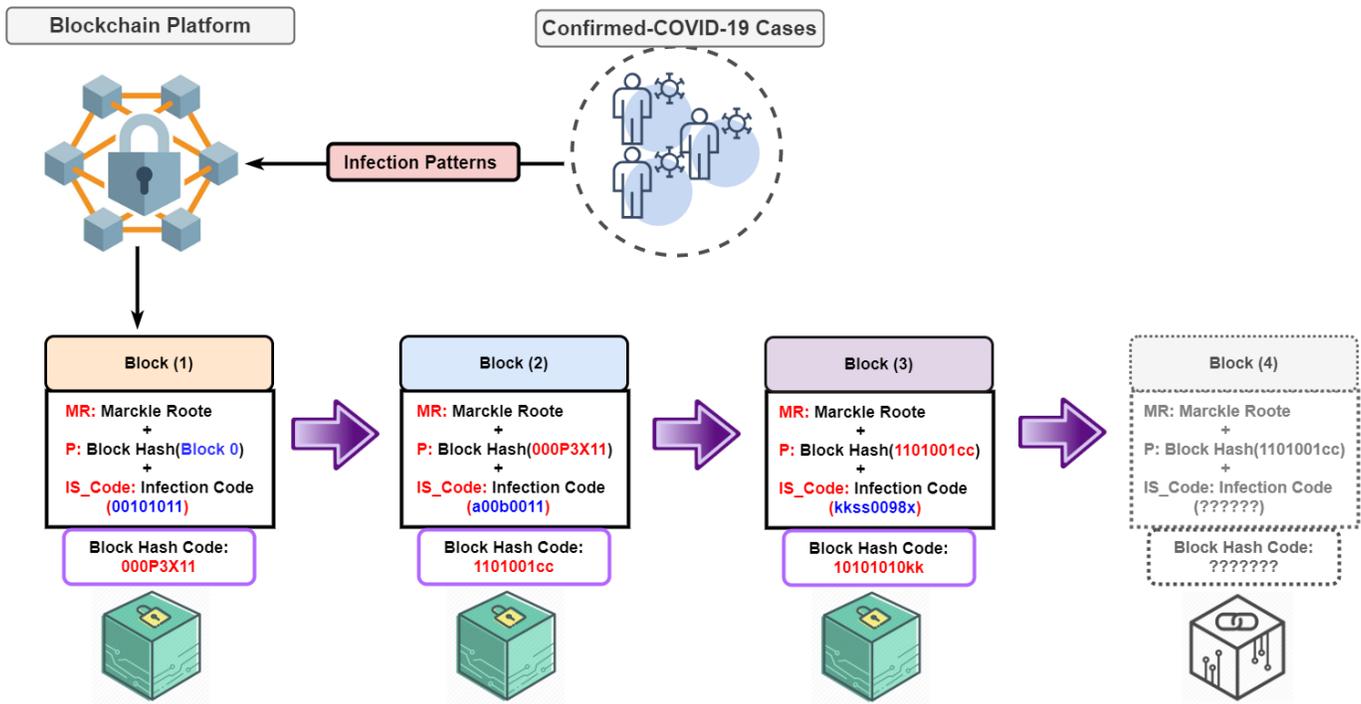

**Figure 6** blockchain design as a sequence of blocks, each block contain a specified number of confirmed COVID-19 cases: A Proposed Model

**Step 4:** when a specific miner succeeded to find the valid hash code of the block first, he then broadcasts this block and its hash code (BHC value) to the remaining miners in the for verifying its validity.

**Step 5:** After broadcasting the new block hash code, the remaining miners verify the validity of the obtained block's hash code. The verification process is done by reversing the hashing function in **equation 1** to check if this hash value leads to the corresponding infection code as in **equation 2.** If the validity of the ne block is confirmed by the reaming miners, all peers will notified that a new block is added to the blockchain. Then, the blockchain state is updated at all nodes with the added block.

$$IS\ Code = Hash^{-1}(MR + P + BHC) \qquad (2)$$

## 2.3 Mass-Surveillance System

Due to the emergent threats that threaten the public security of people and countries, many e-governments (e.g. China, USA, Australia, Germany, Russia, etc) started to track the motion of all citizens and their behaviors by installing many points of Mass-Surveillance System **[16]** in all cities, streets, and public places in order to track and monitor suspicious persons and detect abnormal behaviors as shown in **Figure 7.**

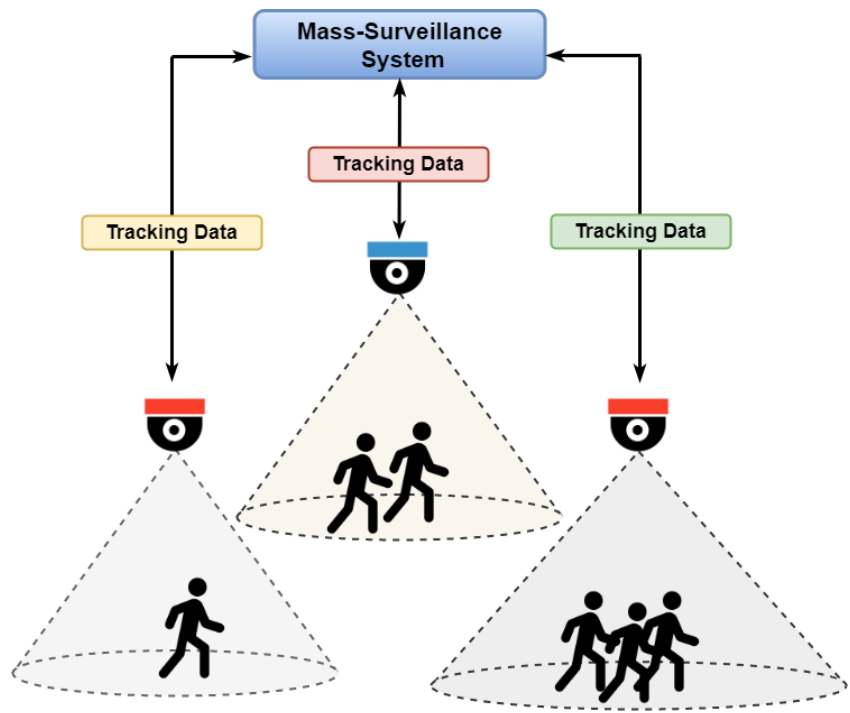

**Figure 7** of   Mass-Surveillance System: A Proposed Model

In this study, we propose to use Mass-Surveillance System for performing two important functions:

1) *Detection and Tracking Function:* using this function, *Mass-Surveillance System* can recognize, detect, and specify the identity of persons for security or health purposes. So, this function can be called to track last behaviors of the confirmed COVID-19 cases for detecting the set of close contact persons who interacted with, and the set of places which COVID-19-patients has been accessed during the last few days. By this function, the governments and hospitals can detect the probable unknown infected cases and risky infected places which frequently have been access by confirmed COVID-19 cases during the last few days.

2) *Blockchain Feedback Function:*  *Mass-Surveillance System* calls this function to send the detected the set of probable infected persons as well as the set of infected places to blockchain which map the set of infected persons and places with the corresponding confirmed COVID-19 cases, then digitize them in the form of *infection patterns* and the follower *infection instances* .

**Figure 8** explains how *Mass-Surveillance System* utilizes *Detection and Tracking Function* and Blockchain *Feedback Function* for recognizing the probable unknown infected persons and places and feedback the blockchain with the answer to the tracking requests. Firstly, blockchain send tracking request to *Mass-Surveillance System* about a specific confirmed-Covid-19 case. Then *Mass-Surveillance System* calls *Blockchain Feedback Function* to refine the tracking request and pass it to *Detection and Tracking Function* to execute it. When the *Detection and Tracking Function* succeeded to recognize and detect all probable infected persons and places according to the handled confirmed-Covid-19 case, the tracking results are passed back to *Blockchain Feedback Function* which then prompt *Mass-Surveillance System* to send the required tracking results (i.e. the set of all probable infected persons and places that are corresponding to the handled confirmed-Covid-19 case) to blockchain platform.

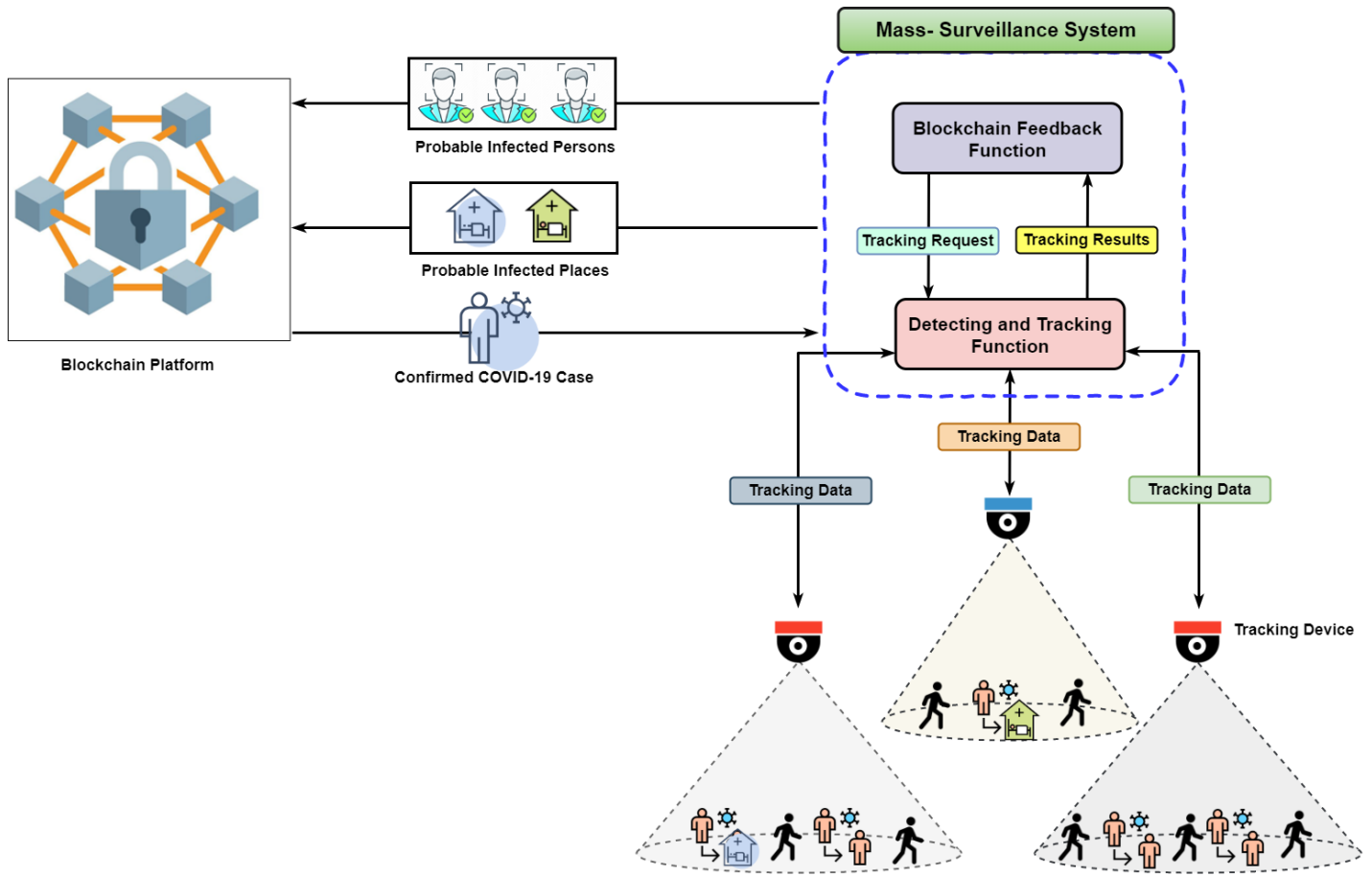

**Figure 8** a proposed design method for Mass-Surveillance System: A Proposed Model

## 2.4 P2P-Mobile Application

A novel design of *P2P-Mobile Application* is proposed also in this framework to embed it with the proposed system. The *P2P-Mobile Application* will provide the citizens with two important services:

1) **Infection Verification Service**. By this mobile service, users are able to receive SMS messages from the blockchain once a novel confirmed COVID-19 case is registered in a specific block in blockchain. The SMS message contain *an infection code* which represent a probable contagion from the registered confirmed COVID-19 case to the user who received an *infection code* in his inbox . For example, suppose a confirmed COVID-19 case has been registered in the blockchain with the *infection pattern, $P/B\text{-}0(1+0)^*101$* , so , the following set of infection codes , { *P/B* 011101, *P/B* 000101, *P/B* 01011101, *P/B* 0000101, *P/B* 0101010101,------} are altogether valid *infection codes* which generated from the pattern , $0(1+0)^*101$.

   Hence, these infection codes (i.e. infection instances) represent all probable infected persons and places which have been contacted with the identified confirmed COVID-19 case. The infection code in the form $P********$ represents probable infected person, while the infection code in the form $B********$ represents probable infected places. So, when a specific person receive an *infection code* from the blockchain, he can use his mobile-P2P mobile application to verify the validity of his infection code  as well as know all details of infection information such as, contagion source (i.e. infection pattern), contagion situation, contagion location, contagion time, etc.

   Each user can verifies his infection code by establishing a connection between his *P2P-mobile application* and the **infection verifier subsystem**. The user can login to the system and enter his received infection code, and then the **infection verifier subsystem** calls the appropriate Finite Automaton machine for verifying the entered code. If the infection code is accepted by the *Finite Automaton Machine*, this mean that the person who hold this code is counted as *a probable infected person* , after that,  the system send him all details of this infection case. **Figure 9** explains how a user can use the infection verification service through his *P2P-mobile application* to verify his *infection code* and be acknowledged with all detailed information about his infection.

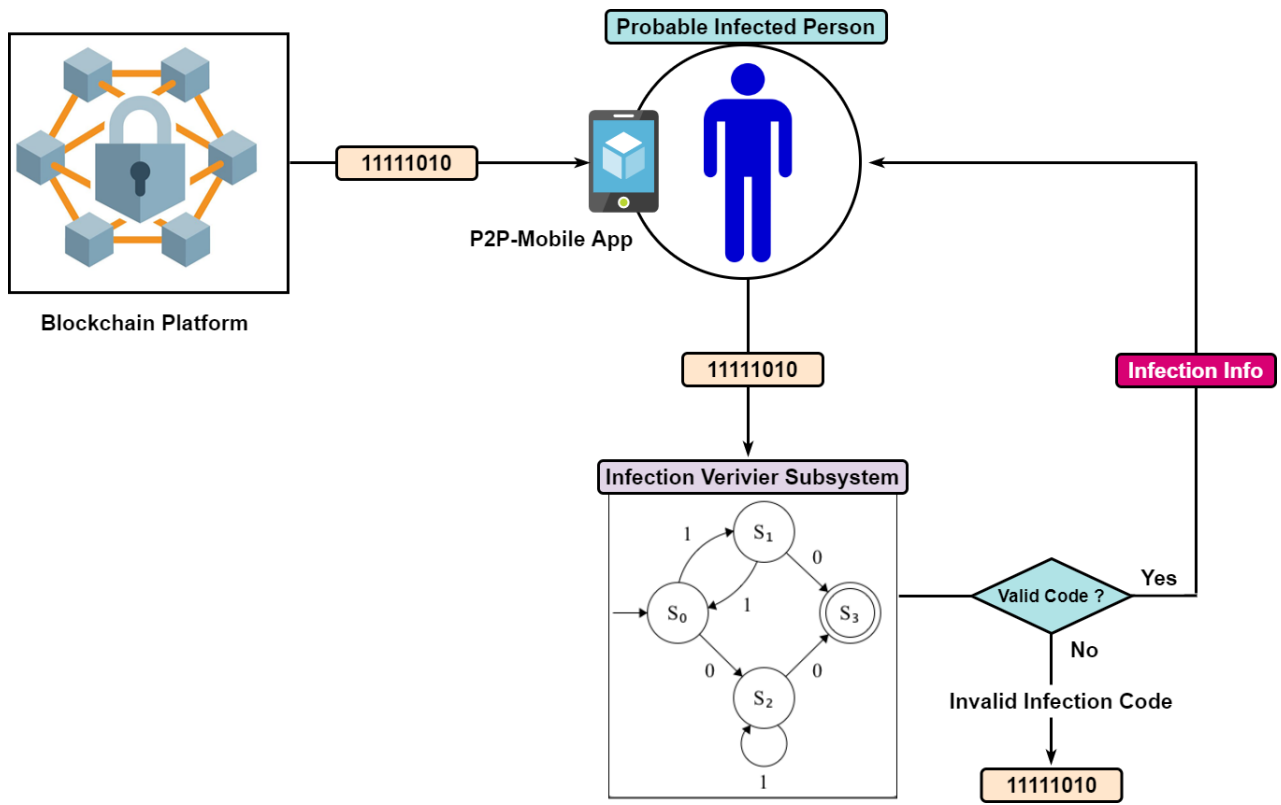

**Figure 9** a proposed Infection Verification Process : A Proposed Model

2) *Infection Risk Estimation and Prediction Service:* This service is the most important one in the P2P-Mobile Application. Using this service, users are able to estimate the probability of infection for them. Moreover, the governments can use this feature to detect the unknown probable infected cases. In addition, two or more peers of citizens can automatically setup a P2P communication through the *P2P-Mobile App* which installed in each one's mobile phone.

So, three important questions rise here regarding using the novel P2P-Mobile Application, the first question is, how users are able to estimate the probability of infection for themselves?. The second question is, how the governments or health authorities are able to detect the probable unknown infected cases or places? Finally, how persons can detect the probable infected cases within different human clusters. We can answer these questions as follows:

*3.4.1 Self estimation of COVID-19 Infection Probability*

One of the interesting services of the proposed P2P-mobile App is the ability of persons to estimate the COVID-19 Infection probability. This goal can be achieved by a hidden communication established between **blockchain platform** and the **P2P-Mobile applications**. when a new confirmed COVID-19 case is registered as a new *infection patterns* and all the corresponding *infection instances* are detected by the **Mass-Surveillance System**, the blockchain platform sends streams of infection codes (in the form SMS messages) to all detected persons (i.e. infected instances) by **Mass-Surveillance System.** Using P2P-Mobile Application, each person can receive many *infection codes* in his inbox from the blockchain. The more confirmed-COVID-19 cases you have contacted with, the more infection codes you receive For example, if you received 10-*infection codes* in your inbox, this mean that you are probably infected from 10-confirmed COVID-19 cases. Using this feature of P2P-Mobile Application, the infection probability of each user can be estimated using **Binomial Distribution Function** as formulated in **equation 3.** This mathematical calculation is done as a hidden routine in the P2P-Mobile Application

$$P(X) = \frac{N!}{X! \times (N-X)!} \times P^X \times Q^{N-X} \qquad (3)$$

Where, $P(X)$ is the infection probability function, $N$ is the total number of *infection codes* in the inbox of each user. $X$ is the number of infection occurs, such that $X = 0,1,2,3,......N$, $P^X$ is the probability of succeeded infection occurrences, and $Q^{N-X}$ is the probability of un-succeeded infection occurrences.

*3.4.2 Detecting Unknown infected Covid-19 Cases.*

Based on the *infection probability* estimations that can be calculated automatically as explained in the previous subsection, then, P2P-*Mobile applications,* send the estimated infection probability values to blockchain platform which can be accessed by governments, health authorities who become able to know the *infection probability rates* of all probable infected COVID-19 cases, hence, they will able to detect unknown infected cases who have the highest infection probability rates.

**Figure 10** explains how infection probability estimations can help the governments and health authorities to detect unknown infected COVID-19 cases.

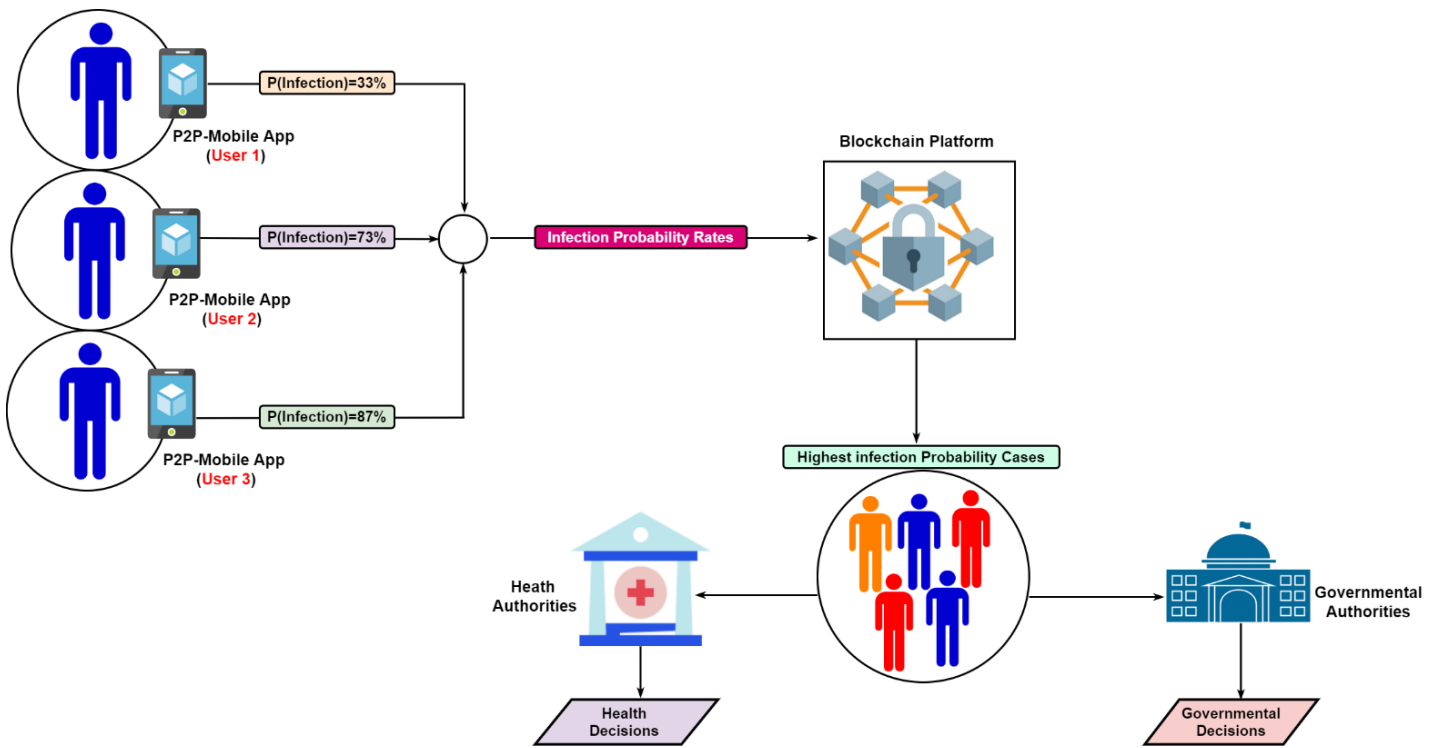

**Figure 10** Detecting unknown Infected Covid-19 cases based Infection probability estimations using the P2P-Mobile Application : A Proposed Model

### 3.4.3 Self-Detection of probable infected cases within Cluster of people

The Peer-to-Peer communication design of the **P2P-Mobile applications**, as well as the decentralized storage design, time stamping, and real time processing advantages of blockchain can enable persons to detect the sources of contagion within a crowdy cluster of people. This goal can be achieved by exchanging the *infection probability rates* between persons using the P2P-*Mobile application* of each person. In other words, when you enter a crowdy cluster of people or a specific person is close to you, your P2P-Mobile Application will automatically receive a *warning message* from the set of all close person(s) through his/their P2P-mobile application. The received warning message contains the identity information of all persons who have infection probability rates. By this way, person can check all received warning *messages* and use his P2P-mobile App to arrange them according to *the highest infection probability rate*, hence, the person become able to take the required precautions procedures against the detected infected cases who have high infection probability rates. **Figure 11**explains how users can detect the probable infected case within a cluster of people using his P2P-Mobile Application

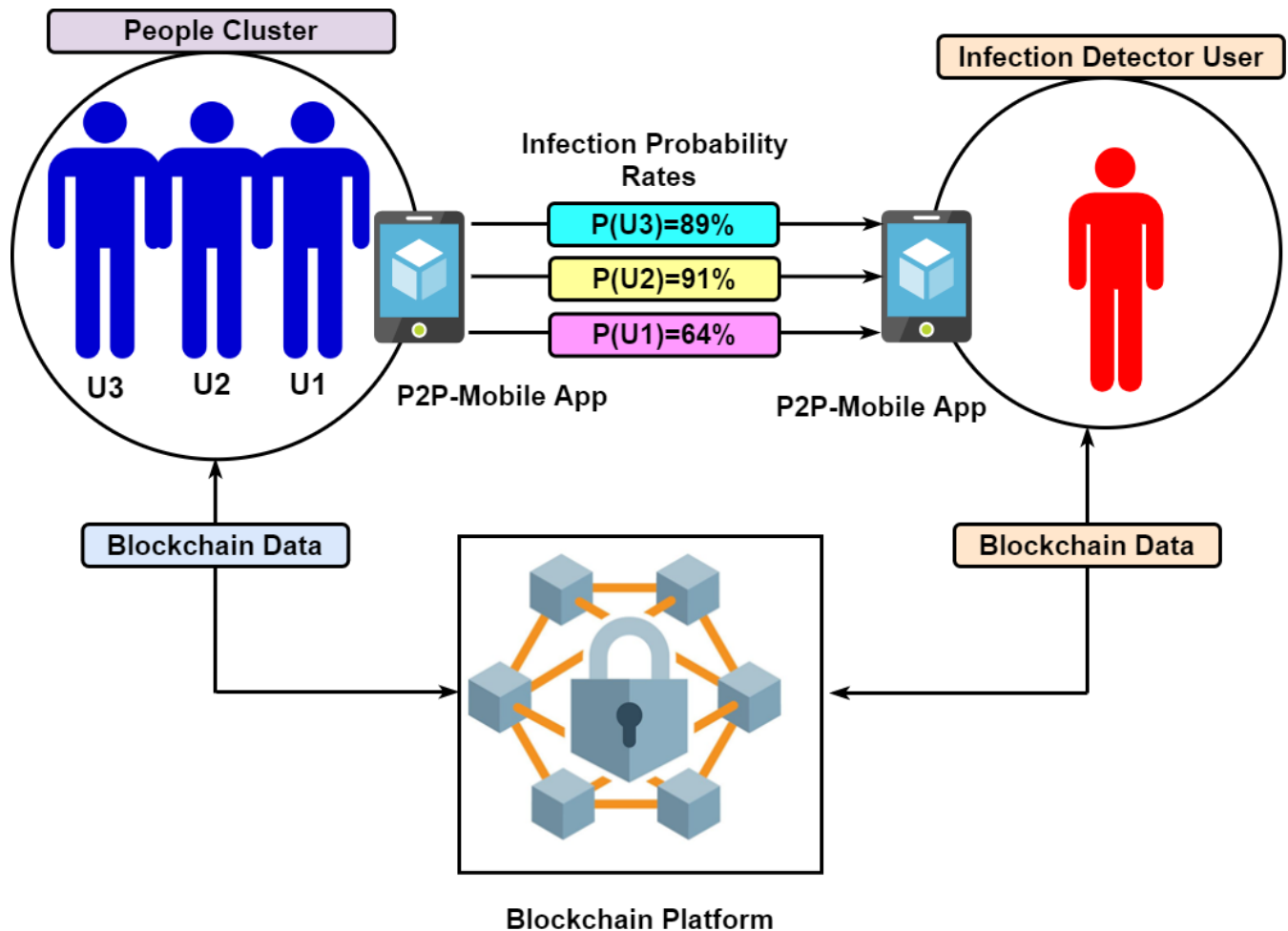

**Figure 11** Self-Detection of probable infected cases within a Cluster of people: A Proposed Model

## 3. Conclusion

The main goal of the current study was to investigate the possibility of developing a novel innovative approach for verifying and detecting the unknown infected cases of coronavirus (COVID-19). This study introduced the first blockchain-framework which can be used for helping governments and health authorities to detect the unknown infected cases and places of coronavirus (COVID-19), as well as automatically estimating and predicting the contagion risks. This new features of the proposed framework is forecasted to be an base support in taking critical decisions by governments, health authorities, and all citizens.

The second aim of this study was to investigate the effects of developing a novel a *blockchain-based P2P-Mobile application design* for helping all citizens to detect the sources of contagion within a crowed cluster of people. Moreover, the novel design of the proposed P2P-Mobile APP will help all citizens to automatically estimate the Infection probability of COVID-19 virus as well as helping the governments and health authorities for detecting unknown infected cases.

This framework is the first comprehensive investigation of utilizing the decentralized design, Peer-to-Peer communication, real time processing, and time stamping advantages of blockchain technology to develop a novel P2P system able to solve many challenges of COVID-19 virus propagation.

The framework is currently being developed and implemented as a new system consists of four components, *Infection Verifier Subsystem*, *Blockchain platform*, *P2P-Mobile Application*, and *Mass-Surveillance System* for investigating the realism and efficiency of the proposed framework.

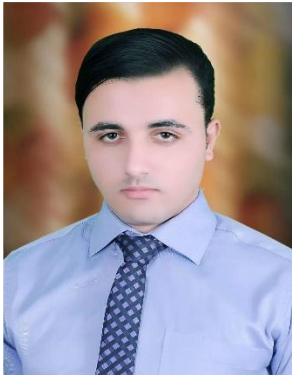

**Dr. Mohamed Torky** is a senior member of the Scientific Research Group in Egypt (SRGE). Dr. Torky works as a Visiting Assistant Professor of Information Technology at faculty of Computer and Information Systems-Islamic University in Medinah - KSA. He worked as Assistant Professor of Computer Science at Higher Institute of Computer Science and Information Systems - Science and Culture City Academy - 6 October City- Giza. Dr. Torky has many publications in prestige journals, proceedings of International conferences, and has two book chapters. His research interests are: Computer and Information Security, Blockchain Technology, Space Science and Satellites, Petri Net- based applications, Graph Theory-based applications, and Automata Theory-based applications.

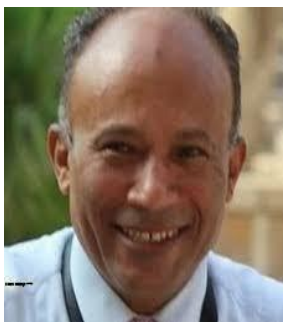

**Dr Aboul Ella Hassanien** is the Founder and Head of the Egyptian Scientific Research Group (SRGE) and a Professor of Information Technology at the Faculty of Computer and Artificial Intelligence, Cairo University. Professor Hassanien has more than 1000 scientific research papers published in prestigious international journals and over 50 books covering such diverse topics as data mining, medical images, intelligent systems, social networks and smart environment. His other research areas include computational intelligence, medical image analysis, space sciences and telemetry mining. Prof. Hassanien won several awards including the Best Researcher of the Youth Award of Astronomy and Geophysics of the National Research Institute, Academy of Scientific

Research (Egypt, 1990). He was also granted a scientific excellence award in humanities from the University of Kuwait for the 2004 Award, and received the superiority of scientific in technology - University Award (Cairo University, 2013). Also He honored in Egypt as the best researcher in Cairo University in 2013. He was also received the Islamic Educational, Scientific and Cultural Organization (ISESCO) prize on Technology (2014) and received the state Award for excellence in engineering sciences 2015. Dr. Hassanien holds the Medal of Sciences and Arts from the first class from President of Egypt "Abdel Fatah Al-Sissy". In 2019, Professor Hassanien received Scopus award for his meritorious research contribution in the field of Computer Science.